\documentclass[aps,twocolumn]{revtex4}
\usepackage[utf8]{inputenc}
\usepackage{graphicx}
\usepackage{amsmath}
\usepackage{mathtools}
\usepackage{amssymb}
\usepackage{color}
\usepackage{amsfonts}
\usepackage{epsf}
\usepackage{epstopdf}
\usepackage{bm}
\usepackage{comment}

\allowdisplaybreaks

\usepackage{hyperref}
\def\equationautorefname~#1\null{Eq.~(#1)\null}
\def\figureautorefname~#1\null{Fig.~#1\null}
\def\tableautorefname~#1\null{Table~#1\null}

\newcommand{\xb}{{\bm x}}

\begin{document}

\title{Universal properties of active membranes}
\author{Francesco Cagnetta,  Viktor \v{S}kult\'{e}ty, Martin R. Evans, and  Davide Marenduzzo}
\affiliation{SUPA, School of Physics and Astronomy, The University of Edinburgh, Edinburgh, EH9 3FD, Scotland, United Kingdom}

\begin{abstract}
We put forward a general field theory for nearly-flat fluid membranes with embedded activators and analyse their critical properties using renormalization group techniques. Depending on the membrane-activator coupling, we find a crossover between \emph{acoustic} and \emph{diffusive} scaling regimes, with mean-field dynamical critical exponents  $z\,{=}\, 1 $ and $2$ respectively. We argue that the acoustic scaling, which is exact in all spatial dimensions, leads to an early-time behaviour which is representative of the spatiotemporal patterns observed at the leading edge of motile cells, such as oscillations superposed on the growth of the membrane width. In the case of mean-field diffusive scaling, one-loop corrections to the mean-field exponents reveal universal behaviour distinct from the Kardar-Parisi-Zhang scaling of passive interfaces and signs of strong-coupling behaviour.

\end{abstract}

\maketitle

Active membranes are fluctuating surfaces which are driven out of thermodynamic equilibrium by the constant action of non-thermal, or {\it active} forces~\cite{ramaswamy2001,turlier2019}. Understanding their behaviour is key to capture the fundamental physics of advancing, or growing, biological interfaces. Indeed, the paradigmatic example of an active membrane is the plasma membrane of eukaryotic cells. Here, activity might enter in many different ways: from intramembranous force sources, such as ion channels~\cite{prost1996,ramaswamy2000} or ATP-consuming membrane proteins~\cite{gov2006,veksler2007,cagnetta1,cagnetta2}; from the coupling with the underlying cytoskeleton~\cite{gov2005,maitra2014, noguchi2021}; from interactions with walls~\cite{prost1998,yasuda2016}. 

Besides having a decisive influence on the membrane fluctuations~\cite{turlier2016}, activity creates a slew of intriguing and non-trivial spatiotemporal patterns, such as lamellipodia or advancing fronts~\cite{insall2009actin,ridley2011life}, longitudinal and transverse waves~\cite{allard2013}, clustering and nonequilibrium phase separation~\cite{gowrishankar2012}. These phenomena occur across a vast variety of biological settings; as such, they may be considered ``universal'' aspects of cell motility~\cite{dobereiner2006, allard2013}. Previous theoretical studies of active membranes, however, have focussed on specific applications or systems of interest~\cite{prost1996,ramaswamy2000,gov2006,veksler2007,cagnetta1,cagnetta2,gov2005,maitra2014,prost1998}: for instance, one set of equations describes the dynamics of curved proteic activators on the membrane~\cite{gov2006, goutaland2020}, while another set studies the instabilities induced by the coupling with the actin cortex~\cite{maitra2014}.  A more general approach, which we adopt here, is to start from a broader framework and ask whether it is possible to identify ``universality classes'' within which to catalogue the dynamics of active membranes.

To address this question we derive equations for the long-wavelength fluctuations of an active membrane with embedded activators. Such activators induce the membrane motion by means of a force normal to the membrane and proportional to their local density. Already at the mean-field level, this model displays two dynamical scaling regimes characterised by distinct values of the dynamical exponent $z$: an {\it acoustic} regime with $z\,{=}\,1$ and a {\it diffusive} regime with $z\,{=}\,2$.  
The acoustic regime potentially describes the aforementioned universal aspects of motile cells dynamics. The diffusive regime contains three different phases: we present minimal models for each, while discussing the resulting scaling laws and their possible relevance in a biophysical context. Corrections to mean-field scaling exponents are calculated within a field-theoretic renormalization group approach (RG)~\cite{cardyScaling,amitFieldTheory,tauberCriticalDynamics,Vasilev04}. This framework, which has proven instrumental in the understanding of equilibrium critical phenomena, has also elucidated universal scaling in active systems made of self-propelled agents~\cite{TonerTu95,toner2005,Chen2015,Cavagna2019,Cavagna2019a,Caballero2020} and growing, albeit passive, interfaces such as the celebrated Kardar-Parisi-Zhang (KPZ) equation~\cite{KPZ,Canet2010}. The details of our one-loop analysis, exact to order $\epsilon=d_c-d$, with $d_c$ the upper critical dimension (ucd) are provided in a companion paper~\cite{Cagnetta2021}.

We begin by describing the system within the \emph{Monge gauge}~\cite{Cai1995}, where the membrane is nearly flat and can be described by specifying its height with respect to a reference plane. The dynamics obey a pair of coupled stochastic differential equations, for the height of the membrane and the density of activators.  We denote with $\bm{x}$ (components $x_a$) the parametrisation of the $d$-dimensional membrane and with $\partial_a$ the partial derivative along the $a$-th spatial direction.
Rather  than the absolute height and density, we consider {\it fluctuations} about a uniform membrane,
with height $\lambda t$ and   density $\rho_0$,  moving with velocity
$\lambda$. Equations for the fields $h(\bm{x},t)$ (height fluctuations) and $\phi(\bm{x},t)$ (excess density) can be derived by separating the forces acting on the membrane-protein system into normal and tangential directions: normal forces affect only the membrane, whereas tangential ones displace both membrane and activators. As a result, to leading order in fields and derivatives~\cite{Cagnetta2021},
\begin{subequations}\label{eq:active-interface-fluctuating-full}
	\begin{align}\label{eq:a-i-f-int-full}
		\partial_t h &= \nu_h \partial_a^2 h +\frac{\lambda}{2}\left(\partial_a h\right)^2 + a_h\phi  + \frac{\alpha}{2}\phi^2  - c_h \partial_a^2 \phi \nonumber\\ & + \sqrt{2 D_h}\xi_n,\\ 
		\label{eq:a-i-f-dens-full}
		\partial_t \phi &= \nu_\phi \partial_a^2 \phi+ \lambda\partial_a\left(\phi\partial_a h\right) + a_\phi\partial_a^2 h - c_\phi \partial_a^4 h \nonumber\\ &+\frac{\kappa}{2}\left[\partial_a^2 (\partial_b h)^2 -2\partial_a\left((\partial_a h)\partial_b^2 h\right)\right] \nonumber \\ &+ \partial_a\left(\sqrt{2 D_\phi}\xi_a\right),
	\end{align}
\end{subequations}
where $\xi_n$ and $\xi_a$ are independent space-time white Gaussian noises.
We stress that \eqref{eq:active-interface-fluctuating-full} contains all leading-order terms that are invariant under reparametrisations of the membrane  in the Monge gauge, such as infinitesimal membrane tilts.

\begin{figure}[t!]
	\centering
	\includegraphics[width=0.4\textwidth]{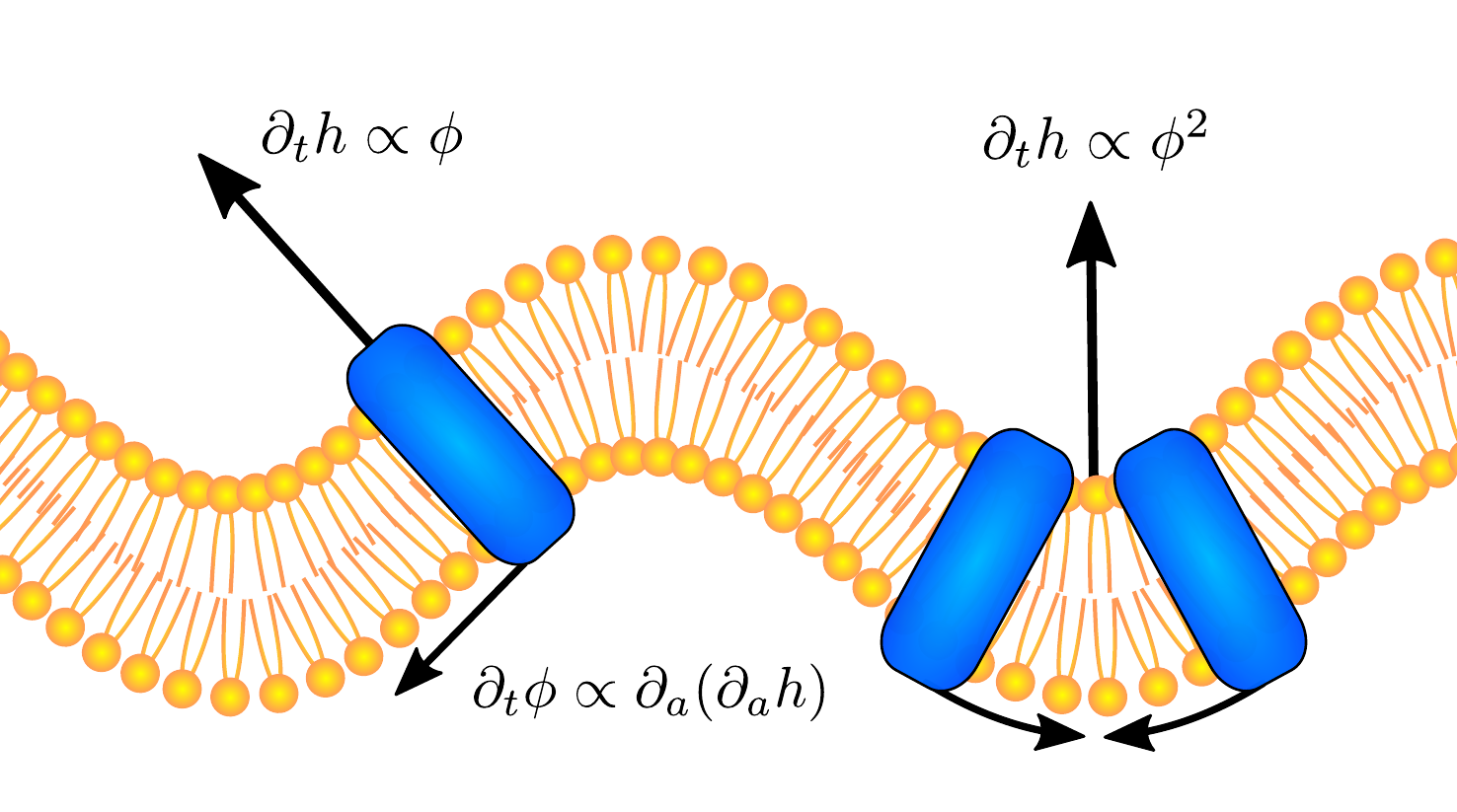}
	\caption{Pictorial representation of the kinematic coupling between vertical motion on the membrane (yellow double-chain) and lateral motion of the activators (blue). The normal force results in vertical displacement of the membrane ($\partial_t h \propto \phi$) and horizontal, slope-dependent displacement of the activators ($\partial_t \phi \propto \partial_a( \partial_a h)$)}
	\label{fig:membrane-activators}
\end{figure}

The first two terms in the height equation~\autoref{eq:a-i-f-int-full}, $\nu_h \partial_a^2 h$ and $\frac{\lambda}{2}(\partial_a h)^2$, represent respectively interfacial tension and a nonlinear correction due to the membrane moving along its own normal---this is analogous to the nonlinear term in the KPZ equation~\cite{KPZ}. The terms $a_h \phi$ and $\alpha \phi^2$ arise from the Taylor expansion of a generic density-dependent driving force $f(\phi)$, whereas the term $-c_h\partial_a^2 \phi$ represents the action of activators which induce a certain curvature on the membrane~\cite{habermann2004bar, vogel2006local, goutaland2020}. The evolution of the activator density~\autoref{eq:a-i-f-dens-full} entails a diffusive term $\nu_\phi \partial_a^2 \phi$ and the advective terms $ \lambda\partial_a\left(\phi\partial_a h\right)$ and $a_\phi\partial_a^2 h$. Such advection by the slope, proportional to the membrane velocity $\lambda$ ($a_\phi\,{=}\,\lambda \rho_0$) arises kinematically because of the vertical motion of the membrane (see ~\autoref{fig:membrane-activators} and~\cite{Cagnetta2021} for details ). The other linear terms, $-c_\phi \partial_a^4 h \equiv -c_\phi \partial_a^2( \partial_b^2 h)$ and $-c_h\partial_a^2 \phi$, form an action-reaction pair which takes into account the possibility that activators possess an intrinsic curvature, which influences the membrane's curvature~\cite{helfrich1988intrinsic, kralj1999free, may2000protein, fosnaric2006influence}. Both $c_\phi$ and $c_h$ are, in principle, proportional to such intrinsic curvature. The nonlinear term proportional to $\kappa$ is the divergence of a nonequilibrium current $\partial_a j_a$: it includes a correction to the curvature coupling term ($\partial_a^2 (\partial_b h)^2$), which causes the activator flow to depend on the local slope as well as the curvature~\cite{krug1997origins}, and a non-gradient contribution, which has a similar origin~\cite{Caballero18}. Thermal fluctuations are represented by the noise terms $\sqrt{2D_{h}\xi_{n}} $ and $ \partial_{a}(\sqrt{2D_{\phi}}\xi_{a}) $. 


We first consider  the deterministic, linearised version of~\autoref{eq:active-interface-fluctuating-full}, which corresponds to the mean-field theory of the problem. Assuming the plane-wave solution $h(\bm{x},t), \phi(\bm{x},t) \propto e^{i \left({\bm{k}}\cdot {\bm{x}}-\omega t\right)}$, implies a dispersion relation linking  mode frequency $ \omega $ and  wavevector $ k $
\begin{equation}\label{eq:dispersion-relation}
-i\omega = -\frac{1}{2} \bigg( \left(\nu_h+\nu_\phi\right)k^2 \pm \sqrt{\left(\nu_h+\nu_\phi\right)^2k^4-4\Delta} \bigg),
\end{equation}
where $ \Delta\equiv\nu_h \nu_\phi k^4 +(a_h + c_h k^2)(a_\phi k^2+ c_\phi k ^4) $. Linear stability requires $ \text{Re}[-i\omega]\,{<}\, 0 $, i.e. $ \nu_{h} + \nu_{\phi} \,{>}\, 0 $ and $ \Delta \,{>}\, 0 $. The latter condition implies an infrared (IR) instability occurring for $ a_{h} a_{\phi} \,{<}\, 0 $~\cite{mahapatra2020}. Such an instability might arise in a membrane driven by an active force linearly proportional to the activator density, in {\em opposition} to a passive homogeneous force. When such instability occurs, the assumption of vanishing slopes $ (\partial_{a} h)^{2}\,{ \ll}\, 1 $ might break down together with the Monge-gauge description. Although exiting the Monge gauge grants access to a variety of stationary shapes~\cite{kabaso2011theoretical, fosnaric2019theoretical, sadhu2021modelling}, in this work we restrict our attention to the stable phase, and therefore set $ a_{h} a_{\phi} \,\geq\, 0 $. Another IR instability, which also restricts the portion of parameter space available to the stable phase, arises for $ \nu_{h} \nu_{\phi} + a_{h}c_{\phi} + a_{\phi}c_{h} \,{<}\, 0 $, the origin of which is the interplay between curvature coupling and active growth \cite{gov2006}. Finally, an ultraviolet (UV) instability $ (k \rightarrow \infty) $ occurs for $ c_{h}c_{\phi}\,{<}\, 0 $. The latter instability does not influence the results of this letter, as the parameters $ c_{h} $ and $ c_{\phi} $ do not appear simultaneously in any of the critical regimes studied below, and is easily cured by including higher-order corrections to diffusion and bending rigidity terms.

In the case $ a_{h} a_{\phi} \,{\geq}\, 0 $, where the system is linearly stable, the nature of the dispersion law \autoref{eq:dispersion-relation} in the IR limit $ k \rightarrow 0 $ depends crucially on the value of $ a_{h}a_{\phi} $. For $ a_{h}a_{\phi} \,{\neq}\, 0 $ the dispersion law is \emph{acoustic}, $ \omega \simeq \pm (a_{h}a_{\phi})^{1/2} k$ i.e. $\omega\sim k^z$ with $z\,{=}\,1$, whereas for $ a_{h}a_{\phi}\,{=}\, 0 $ it is \emph{diffusive}, $ \omega \sim k^{2} $ thus $z\,{=}\,2$. This leads to two distinct scaling regimes depending on whether $ a_{h}a_{\phi}\,{=}\, 0 $ or not, as we now justify within the Wilsonian RG approach. This approach is well suited to describe the \emph{crossover} between the two regimes at the linear level: the detailed field-theoretic RG analysis of the full model and the calculation of one-loop corrections to scaling exponents can be found in~\cite{Cagnetta2021}. 
 
By rescaling spatial and temporal variables $ (\xb,t) \rightarrow (\xb/b, t/b^{z}) $, and the parameters and fields  according to $ \psi \rightarrow b^{y_{\psi}} \psi $, we obtain the following RG equations for the coefficients of the lowest-order terms of~\autoref{eq:active-interface-fluctuating-full},
\begin{subequations}\label{eq:RG}
	\begin{align}
		\partial_{l} a_{h} &= ( z - y_{\phi} + y_{h} \dots ) a_{h}, \label{eq:RG.1} \\
		\partial_{l} a_{\phi} &= ( z - 2 + y_{\phi} - y_h + \dots ) a_{\phi}, \label{eq:RG.2} \\
		\partial_{l} \nu_{h} &= ( z - 2 + \dots ) \nu_{h}, \label{eq:RG.3} \\
		\partial_{l} \nu_{\phi} &= ( z - 2 + \dots ) \nu_{\phi}, \label{eq:RG.4} \\
		\partial_{l} D_{h} &= ( z - d + 2 y_{h} + \dots ) D_{h}, \label{eq:RG.5} \\
		\partial_{l} D_{\phi} &= ( z - d - 2 + 2 y_{\phi} + \dots ) D_{\phi}, \label{eq:RG.6} 
	\end{align}
\end{subequations}
where $ l\,{=}\, \ln b $. The dots denote corrections coming from nonlinearities which vanish in mean-field. Scale invariance is found at the fixed points of the RG equations, Eqs.~(\ref{eq:RG.1}--\ref{eq:RG.6}),  which  depend on the values of $ a_{h} $ and $ a_{\phi} $. For instance, for $ a_{\phi}, a_{h} \,{>}\, 0 $, for which the dynamical exponent is $z\,{=}\,1$, Eqs.~(\ref{eq:RG.3},\ref{eq:RG.4}) show that any fixed point requires the parameters $\nu_h$ and $\nu_\phi$ to be $0$. The situation is fundamentally different for $ a_{h}a_{\phi} \,{=}\, 0 $, where $z\,{=}\,2$. For example, if both $ a_{h} $ and $ a_{\phi} $ vanish in the starting model, the right-hand sides of Eqs.~(\ref{eq:RG.1},\ref{eq:RG.2}) vanish identically, and the right-hand sides of Eqs.~(\ref{eq:RG.3},\ref{eq:RG.4}) vanishes because $z\,{=}\,2$. Therefore $\nu_h$ and $\nu_{\phi}$ are now marginal in the RG sense. In other words, the presence or absence of the linear `active' terms involving $ a_{h} $ and $ a_{\phi} $ in \autoref{eq:active-interface-fluctuating-full} is crucial to determining the relevant dynamic exponent.

\begin{figure}[t!]
	\centering
	\includegraphics[width=0.5\textwidth]{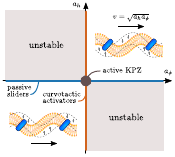}
	\caption{Phase diagram of the active interface model in the $a_h-a_\phi$ plane. The linearly-unstable region (second and fourth quadrant) cannot be accessed within our framework. All membranes with strictly positive $a_h a_\phi$ can be characterised with the same set of scaling exponents, corresponding to the generic active membrane universality class. As shown by the cartoon, the dynamics in this region of the parameter space is dominated by coupled density and height waves having speed $v\propto\sqrt{a_h a_\phi}$. Tuning the wave speed to zero leads to distinct scaling regimes whose properties are discussed in the text.}
	\label{fig:phase-diagram}
\end{figure}

The projection of the phase diagram on the $a_h$-$a_\phi$ plane is shown in~\autoref{fig:phase-diagram}. We now describe the scaling properties of the various regimes and their relevance to active membranes physics.

\textbf{Generic active membrane.} If $a_\phi a_h >0 $, the natural dynamic scaling is \emph{acoustic}, with $ z = 1 $. As this is the typical case, we call this the {\em generic active membrane} phase. In this regime, the system behaviour is controlled by kinematic waves~\cite{ramaswamy2000, maitra2014, cagnetta1} with velocity $v\propto \sqrt{a_h a_\phi}$. As the parameters $a_h$ and $a_\phi$ are connected to the average density of activators and the speed of the membrane motion \cite{Cagnetta2021}, this observation provides a relation between the vertical speed of the membrane and the lateral speed of membrane waves which is amenable to experimental testing. Rescaling under~\autoref{eq:RG} leads to the following linear stochastic model, which describes the acoustic regime
\begin{subequations}\label{eq:active-interface-GAM-intro}
	\begin{align}\label{eq:a-i-GAM-intro-int}
		\partial_t h &=  a_h\phi + \sqrt{2 D_h}\xi_n, \\ 
		\label{eq:a-i-GAM-intro-dens}
		\partial_t \phi &= a_\phi\partial_a^2 h + \partial_a\left(\sqrt{2 D_\phi}\xi_a\right),
	\end{align}
\end{subequations}
We  stress that, although both viscosities $ \nu_{h} $ and $ \nu_{\phi} $ have disappeared, equations (\ref{eq:active-interface-GAM-intro}) should be interpreted as
as an inviscid limit $\nu_h\,{=}\,\nu_\phi\,{\to}\,0^+$.
In fact, in the absence of viscosity (\ref{eq:active-interface-GAM-intro}) would not admit a stationary state and the height and density fields would grow unbounded~\cite{Cagnetta2021}. In other words  the  viscous coefficients $\nu_h$,$\nu_\phi$
in (\ref{eq:active-interface-fluctuating-full}  are \emph{dangerously} irrelevant parameters.

Remarkably, as far as the acoustic regime is concerned, the other linear terms of~\autoref{eq:active-interface-fluctuating-full}, as well as  the nonlinear couplings, all turn out to be RG-irrelevant. 
\begin{table}
	\def\arraystretch{1.3}
	\centering
	\begin{tabular}{ p{0.28\textwidth}  c  c  c }
		\hline \hline
		Model & $ \ z \ $  & $ y_{h} $ & $ y_{\phi} $
		\\  \hline 
		Generic Active Membrane & $ 1 $ & $ (d-1)/2 $ & $ (d+1)/2 $ 
		\\
		Active KPZ (MF) & $ 2 $ & $ (d-2)/2 $ & $ d/2 $ 
		\\
		Active KPZ $ (\lambda^{*} = 0) $ & $ 2 $ & $ d-2 $ & $ d/2 $
		\\
		Curvotactic activators (MF) & $ 2 $ & $ (d-4)/2 $ & $ d/2 $ 
		\\
		Passive sliders (MF) & $ 2 $ & $ (d-2)/2 $ & $ (d-2)/2 $ 
		\\\hline \hline
	\end{tabular}
	\caption{ Universal scaling exponents for $a_h a_{\phi}\geq0$ (generic active membrane) and for $a_h=a_{\phi}=0$ (active KPZ) . MF denotes mean-field exponents. The generic active membrane exponents are exact in any dimension, where those of active KPZ at the $\lambda\,{=}\,0$ fixed point represent one-loop results below $d\,{=}\,2$.}
	\label{tab:canon_CKPZ86}
\end{table}
The fact that all nonlinearities are irrelevant implies that the scaling exponents derived within the mean-field approximation are \emph{exact} in all dimensions. However, the presence of vanishingly small viscosities  in  \eqref{eq:active-interface-fluctuating-full}  implies a steady-state behaviour which is ultimately controlled by a diffusive scaling regime. The acoustic regime is nonetheless relevant for the scaling of all phenomena occuring across time- and lengthscales which are linearly proportional to each other. The exponents are summarised in the first row of~\autoref{tab:canon_CKPZ86} and we now discuss some implications in the context of relaxation from a homogeneous initial condition with $h\,{=}\,\phi\,{=}\,0$.

First, $z\,{=}\,1$ is the natural exponent to describe the waves of protrusion and activator density seen at the leading edge of motile cells~\cite{gowrishankar2012, allard2013}. Secondly, the  exponent $y_h$ controls the scaling of height-height correlations. In the field of kinetic roughening~\cite{krug1991}, one is often concerned with the development in time of the height structure factor $S_h(\bm{k}, t) \propto \left\langle h(\bm{k},t)h(-\bm{k},t) \right\rangle$ after preparing the interface in a flat initial condition. The height structure factor obeys the scaling relation,
\begin{equation}\label{eq:family-vicsek}
 S_h(\bm{k}, t) \simeq k^{2y_h-d}\mathcal{S}_h(k^zt)
\end{equation}
for small wavevectors $k\sim 1/L$ with $L$ the system size. In the acoustic regime (times of order $t\sim 1/k$)
$ S_h(\bm{k}, t) \simeq  k^{-1}\mathcal{S}_{h,1}(kt)$. The presence of a small viscosity $\nu_h$ causes a crossover to a diffusive regime when $t\sim (\nu_hk^2)^{-1}$, with $ S_h(\bm{k}, t) \simeq  k^{-2}\mathcal{S}_{h,2}(k^2t)$ which is the scaling observed in the Edwards-Wilkinson (EW)
equation for equilibrium interface fluctuations  \cite{EW} (exponents as in the second row of~\autoref{tab:canon_CKPZ86}). The acoustic component of $S_h(\bm{k},t)$ influences macroscopic observables such as the squared \emph{width} of the interface, i.e. the variance of the height profile $w^2=\sum_{\bm{k}\neq \bm{0}} S_h(\bm{k},t)$, implying width oscillations with period diverging with $L$ superposed on the usual EW scaling. Such oscillations were observed numerically in~\cite{cagnetta1} for a lattice model which falls in the generic-active-membrane scenario. Similarly, the exponent $y_\phi$ controls the early-time scaling of the density-density correlations, with structure factor in the acoustic regime
$ S_\phi(\bm{k}, t) \simeq  k\mathcal{S}_{\phi,1}(kt)$.
It is worth noting the vanishing of the structure factor for $k\to 0$, which is typical of \emph{hyperuniform} states~\cite{torquato2003}. Also $S_\phi$, because of the small viscosity $\nu_\phi$, crosses over to a diffusive structure factor $S_\phi(\bm{k}, t) \simeq  \mathcal{S}_{\phi,2}(k^2t)$ when $\sim (\nu_\phi k^2)^{-1}$.


\begin{figure}[t!]
	\begin{center}
		\begin{tabular}{cc}
			\includegraphics[width=\columnwidth]{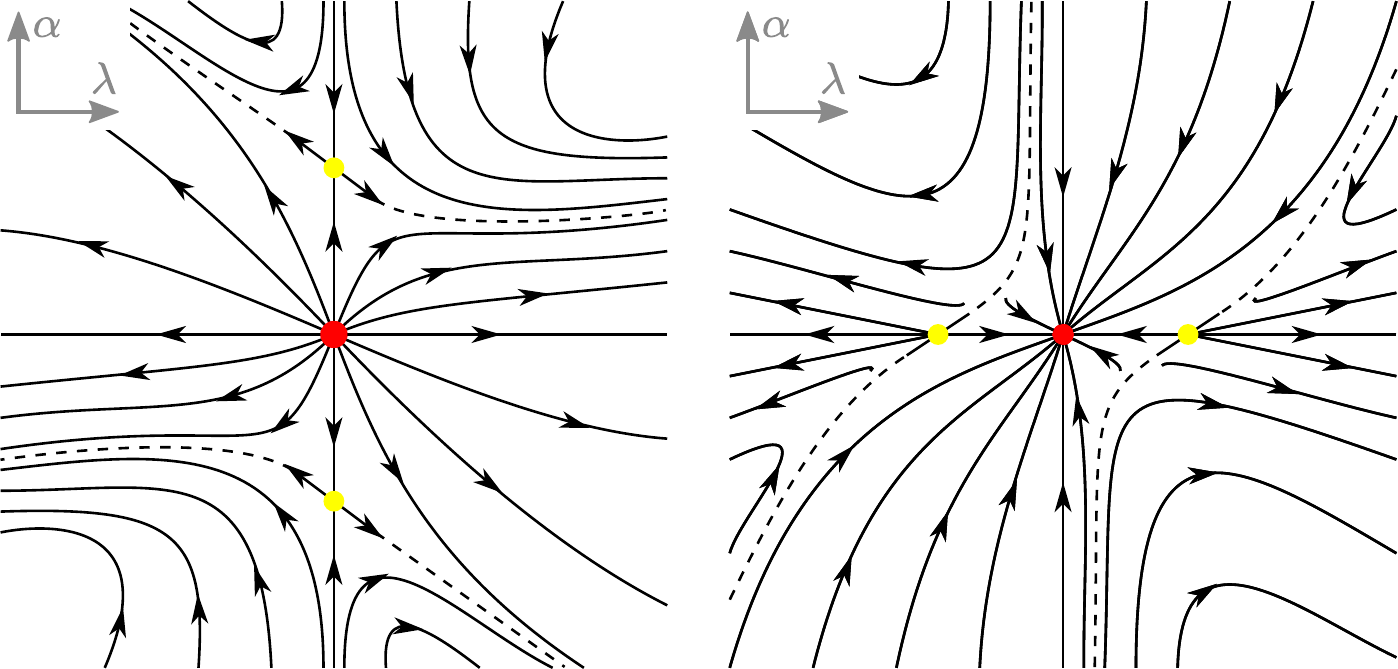}
		\end{tabular}
		\caption{RG flow of the active KPZ model \eqref{eq:active-interface-AKPZ-intro} in the $\alpha$-$\lambda$ plane. For $d\,{<}\,2$ (a), there is a perturbative fixed point on the $\lambda\,{=}\,0$ line, marked by the yellow dot. For $d\,{>}\,2$ (b) the flow converges onto the Gaussian fixed point along the $\alpha$ direction. The yellow dot on the $\alpha\,{=}\,0$ line marks the IR-unstable fixed point associated with the roughening transition of the KPZ equation.}
		\label{fig:phase-diagram-KPZ}
	\end{center}
\end{figure}

\textbf{Active KPZ.}  At the point $a_h\,{=}\,a_\phi\,{=}\,0$ (see phase diagram in~\autoref{fig:phase-diagram}), there is \emph{diffusive} mean-field scaling in the thermodynamic limit with $z\,{=}\,2$. After simple rescaling, the evolution of the membrane and activator fields is given by the following coupled equations
\begin{subequations}\label{eq:active-interface-AKPZ-intro}
	\begin{align}
		\partial_t h &= \nu_h \partial_a^2 h + \frac{\lambda}{2}(\partial_a h)^2 + \frac{\alpha}{2}\phi^2 + \sqrt{ \nu_{h} }\xi_n;\\
		\partial_t \phi &=  \nu_\phi \partial_a^2 \phi + \lambda\partial_a(\phi\partial_a h) + \partial_a\left( \sqrt{ \nu_{\phi} }\xi_a\right),
\end{align}\end{subequations}
\autoref{eq:active-interface-AKPZ-intro} constitutes what we call the \emph{active} KPZ model, which is the minimal model for $a_h\,{=}\,a_\phi\,{=}\,0$. Activity comes from the $\frac{\alpha}{2}\phi^2$ term, which may arise physically when clustering of activators is required to stimulate membrane growth~\cite{zakine2018}.
Our RG analysis shows this nonlinear growth term to be marginal, thus influencing the critical properties of the model, only for $a_h\,{=}\,a_\phi\,{=}\,0$ and otherwise irrelevant. The ucd of the model is $d_c\,{=}\,2$.

Just above the critical dimension, a perturbative unstable fixed point at $\lambda\,{=}\,\lambda_c$ (yellow dot in the right panel of~\autoref{fig:phase-diagram-KPZ}) marks the famous roughening transition of the KPZ equation~\cite{frey1994aa}. In the $\alpha$-$\lambda$ plane, the KPZ fixed point has an additional marginal direction, implying that $\lambda_c$ depends linearly on $\alpha$ for $\alpha$ small. Inside the region bounded by the two dashed lines, the scaling exponents attain their mean-field values $y_h\,{=}\,(d-2)/2$ and $y_\phi\,{=}\,d/2$, which are the typical values at the Gaussian fixed-point for a non-conserved and conserved order parameter, respectively. Outside this region the scaling is controlled by a non-perturbative fixed point~\cite{Canet2010}. 
Below the ucd  $d_c\,{=}\,2$, the $ \alpha $-term generates a novel perturbative fixed point with $\lambda\,{=}\,0$~\cite{Cagnetta2021} (yellow dot in the left panel of~\autoref{fig:phase-diagram-KPZ}). As $\lambda\,{=}\,0$, this fixed point describes a membrane with no net vertical motion, whose fluctuations are controlled by the distribution of activators through the $\alpha$-term. One-loop corrections to the scaling exponents are listed in the second row of~\autoref{tab:canon_CKPZ86}. At  lowest order in $d{-}2$, the dynamic and density exponents retain their mean-field values $y_\phi\,{=}\,d/2$, and $z\,{=}\,2$. The exponent $y_h$, instead, increases and equals $y_h\,{=}\,d-2$. The physical implication is that active fluctuations generate a much rougher interface than thermal fluctuations in $d\,{=}\,1$, with the square width $w^2\sim L^2$ instead of $L$.

\textbf{Curvotactic activators.} Along the line $a_\phi\,{=}\,0$, 
 scale invariant behaviour is characterised by $y_h\,{=}\,(d-4)/2$ and $y_\phi\,{=}\,d/2$ in mean-field. Consequently, $ \nu_{h},\nu_{\phi},a_{h} $ and $ D_{\phi} $ are marginal in the RG equations~\autoref{eq:RG} and need to be included in the minimal model, whereas $D_h$ flows to zero. The minimal equations of motion have then the following form
\begin{subequations}\label{eq:active-interface-diffusive-curvotactic-intro}
	\begin{align}\label{eq:a-i-diff-i-curvotactic-intro}
		\partial_t h &= \nu_h \partial_a^2 h +\frac{\lambda}{2}\left(\partial_a h\right)^2 + a_h \phi,\\ 
		\label{eq:a-i-diff-dens-curvotactic-intro}
		\partial_t \phi &= \nu_\phi \partial_a^2 \phi + \lambda\partial_a\left(\phi\partial_a h\right) - c_\phi \partial_a^4 h + \partial_a \big(\sqrt{ \nu_\phi}\xi_a\big)  \nonumber\\ & + \frac{\kappa}{2}\left[\partial_a^2 (\partial_b h)^2 -2\partial_a\left((\partial_a h)\partial_b^2 h\right)\right].
	\end{align}
\end{subequations}
The most significant  difference with respect to the cases discussed above is the irrelevance of the noise term in the height equation. The fluctuations of the height are instead controlled by the noise acting on the activator density, via the active term $a_h\phi$. Such a noise transfer effect results in the interfacial width scaling with system size with an exponent equal to $-y_h\,{=}\,(4-d)/2$, resulting in more pronounced roughening with respect to a KPZ interface at the mean-field level. However, around the ucd $d_c\,{=}\,4$, the presence of the linear curvotactic term $\propto c_\phi$ and the nonequilibrium current $\propto \kappa$ changes the nature of the RG flow~\cite{Cagnetta2021}. Our analysis shows that, in the one-loop approximation, a new repulsive fixed point emerges below $d_c$, suggesting that the scaling properties of the model might be governed by a non-perturbative fixed point. 

\textbf{Passive sliders.} Finally, on the line $a_h\,{=}\,0$, 
 scale invariant behaviour is characterised by $y_h\,{=}\, y_\phi \,{=}\, (d-2)/2$, thus $\nu_h$, $\nu_\phi$, $a_\phi$ and $D_h$ are marginal and $D_\phi$ is irrelevant. The minimal equations of motion for this model are
\begin{subequations}\label{eq:active-interface-diffusive-passivesliders-intro}
	\begin{align}
		\partial_t h &= \nu_h\partial_a^2 h +\frac{\lambda}{2}\left(\partial_a h\right)^2 -c_h\partial_a^2 \phi+\sqrt{\nu_h}\xi_n,\\ 
		\label{eq:a-i-diff-dens-passivesliders-intro}
		\partial_t \phi &= \nu_\phi \partial_a^2 \phi+ \lambda\partial_a\left(\phi\partial_a h\right) +  a_\phi\partial_a^2 h.
	\end{align}
\end{subequations}
In the special case where  $c_h=0$ the activators do not influence the interface dynamics and the equations  describe a system of passive particles which are advected by the slopes of a fluctuating KPZ interface---the passive sliders model studied in~\cite{das2000,drossel2002,nagar2005}. The passive sliders phase has `naive' ucd $d_c=2$, above which mean-field predictions should hold. Because of a noise-transfer mechanism analogous to that seen in the curvotactic activators model, the mean-field density scaling exponent is much smaller than in the active KPZ model, $y_\phi=(d-2)/2$, resulting in pronounced number fluctuations. This model is, however, non-renormalizable as the RG procedure generates and infinite number of marginal terms~\cite{Cagnetta2021}, and standard methods cannot be used---this feature may be considered a sign of non-universal behaviour \cite{antonov2017}. 

In summary, we have presented and studied a general model for active membranes. The model describes a nearly-flat membrane, populated with an ensemble of activators which stimulate its motion. The most important result of our detailed RG analysis is the crossover between two distinct scaling regimes, caused by structure of the membrane height and activator density interaction. In the presence of linear density-height couplings $ a_{h}a_{\phi} > 0 $, the system belongs to the ``generic active membrane'' universality class, characterised by a linear dispersion law.  Such a scenario is relevant to the physics of lamellipodia formed at the leading edge of eukaryotic cells, where transverse waves travel along the advancing membrane~\cite{allard2013}. Our analysis links the emergence of such waves to particular scaling laws for the fluctuations of membrane height and activator density, which are testable in scattering experiments. 

In the case where membrane growth requires a nonlinear interaction between activators, or when the membrane is kept stationary, the dispersion law \eqref{eq:dispersion-relation} is quadratic in $k$ in mean-field. Here, three different scaling limits are accessible depending on how the condition $ a_{h}a_{\phi}\,{=}\, 0 $ is realised. The active force $ \frac{\alpha}{2} \phi^{2} $ in the height equation emerges in the absence of any linear interaction, and the resulting equations describe the ``active KPZ'' model. A novel perturbative fixed point exists for non-advancing interfaces, with a roughness exponent larger than that of KPZ and other passive interfaces~\cite{KPZ}. In the advancing-interface case the RG flow shows a runaway solution, indicative of a strong coupling behaviour similarly to passive KPZ interfaces~\cite{Canet2010}. Different diffusive regimes, the ``curvotactic activators'' and ``passive sliders'' phases, appear when $ a_{\phi}\,{=}\,0 $ or $ a_{h} \,{=}\, 0 $ respectively, both describing different physical scenarios. Simplified versions of these models appeared in~\cite{gov2006} and~\cite{das2000,drossel2002,nagar2005} on a phenomenological basis, which demonstrates that our field-theoretic approach also leads to the unification of previous studies of active membranes. 


FC and V\v{S} contributed equally to this work. The authors would like to thank Juha Honkonen, Ananyo Maitra and Sriram Ramaswamy for many illuminating discussions. FC and DM acknowledge funding from ERC (Consolidator Grant THREEDCELLPHYSICS, 648050). V\v{S} acknowledges funding from EPSRC Grant No. EP/L015110/1 under a studentship.

\bibliographystyle{apsrev4-1}
\bibliography{ActiveMembrane,FieldTheory}

\end{document}